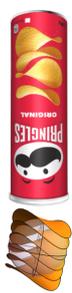

# Quantum Potato Chips

**Nikolay Murzin**
nikm@wolfram.com
Wolfram Institute, USA
Wolfram Research, USA

**Bruno Tenorio**
brunot@wolfram.com
Wolfram Research South
America, Peru

**Sebastian Rodriguez**
srodriguez@wolfram.com
Wolfram Research South
America, Peru

**John McNally**
jmcnally@wolfram.com
Wolfram Research, USA

**Mohammad Bahrami**
mbahrami@wolfram.com
Wolfram Research, USA

2024-11-01

## Abstract

We examine qubit states under symmetric informationally-complete measurements, representing state vectors as probability 4-vectors within a 3-simplex in $\mathbb{R}^4$. Using geometric transformations, this 3-simplex is mapped to a tetrahedron in $\mathbb{R}^3$. A specific surface within this tetrahedron allows for the separation of probability vectors into two disjoint 1-simplices. The intersection of this surface with the insphere identifies a "quantum potato chip" region, where probability 4-vectors reduce to two binary classical variables. States within this region can be fully reconstructed using only two given projective measurements, a feature not found elsewhere in the state space.

## 1. Introduction

A qubit in a normalized quantum state can be expressed in the basis $\{\mathbb{I}, \sigma_x, \sigma_y, \sigma_z\}$ as a 4-vector $\frac{1}{2}\{1, x, y, z\}$, with $\{x, y, z\} \in \mathbb{R}^3$. Since the first component of the 4-vector is fixed, the state can be fully described by the Bloch vector $\vec{r} = \{x, y, z\}$ in $\mathbb{R}^3$. The condition $|\vec{r}| \leq 1$ ensures that the density matrix is positive semi-definite, corresponding to a *physical* state. Alternatively, in an appropriate basis, the qubit state can be represented as a probability 4-vector $\{p_1, p_2, p_3, p_4\}$ with $\sum_i p_i = 1$ and $p_i \geq 0$, known as the probability phase-space representation. Symmetric Informationally Complete Positive Operator-Valued Measures (SIC-POVMs) are optimal for such representations due to their symmetry and informational completeness [1–6]. That said, the phase-space representation has historically been associated with bases such as Wigner, Wootters, and Gell-Mann [7–9], which require a quasi-probability approach (i.e., $\sum_i p_i = 1$ but some $p_i$ can be negative). In this paper, we focus primarily on POVMs, though we also address the quasi-probability case.

For a qubit state in a SIC-POVM, the probability vector is confined within a simplex embedded in $\mathbb{R}^4$. Through appropriate geometric transformations, this 4D object can be linearly projected onto a tetrahedron in $\mathbb{R}^3$. Two key questions arise. From probability theory perspective, among all the points in the tetrahedron, which ones can be reduced to a disjoint pair of 1-simplices (two line segments), representing a separable probability distribution of uncorrelated binary random variables? From quantum theory perspective, which points within the tetrahedron correspond to *physical* quantum states? With answers to these questions in hand, we will be able to find quantum states that correspond to two disjoint probability distributions. This is the core idea of quantum potato chips that we discuss in this paper.

We will show that for states within the quantum potato chips, it is possible to reconstruct the entire state from only two independent projective measurements. This is possible because their probability vectors can be reduced into two disjoint separable ones. This allows straightforward embedding of classical probability structures within quantum state representations, offering a novel intersection between classical and quantum probabilistic frameworks. An immediate consequence of the existence of these

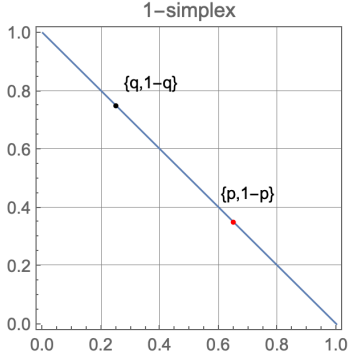 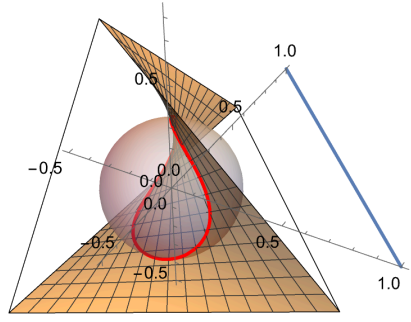 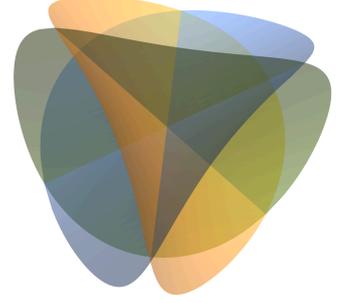

(a) A 1-simplex as the lowest dimensional probability space

(b) Intersection of tetrahedron and surface of two 1-simplex probability spaces

(c) Three quantum potato chips

Figure 1: (a) A 1-simplex defined by the points $\{1,0\}$, $\{0,1\}$ (solid blue line). (b) the blue solid line represents 1-simplex (same as in Figure 1a). The surface, described by Eq.(5), lies entirely within the 2-simplex (a tetrahedron).The solid red line represents the intersection of this surface with the tetrahedron's insphere. Only the points within the insphere correspond to valid physical quantum states. (c) quantum potato chips are defined in Eq.(5), Eq.(8), and Eq.(9), and parametrized by Eq.(7). With one such surface in hand, one can find the other two through proper permutation of variables.

states is that any classical problem with two binary variables can be mapped into qubits. Future work will examine applications and foundational implications in more detail.

**Note for the readers**: All visualizations and some formulas in this paper have corresponding Wolfram Mathematica code, which can be found at the end. These are accessible by clicking on equations or figure captions, if applicable. Here is the complete Wolfram notebook: https://wolfr.am/QPC

## 2. Generating Quantum Potato Chips

### 2.1. SIC-POVM Case

In an Informationally-Complete POVM (IC-POVM), a qubit state is fully described by a probability 4-vector $\{p_1, p_2, p_3, p_4\}$ with $\sum_i p_i = 1$ and $p_i \geq 0$. The space of probability vectors is the unit standard simplex in $\mathbb{R}^4$, spanned by points $\{1,0,0,0\}$, $\{0,1,0,0\}$, $\{0,0,1,0\}$, and $\{0,0,0,1\}$. This is a region with the embedding dimension 4 and the geometric dimension 3. Therefore, one can reduce the embedding dimension to 3. Consider the following rotation matrix:

$$U_{\text{rot}}(\theta) = \begin{pmatrix} \cos\theta & \frac{\sin\theta}{\sqrt{3}} & \frac{\sin\theta}{\sqrt{3}} & \frac{\sin\theta}{\sqrt{3}} \\ -\frac{\sin\theta}{\sqrt{3}} & \frac{1}{3}(\cos\theta+2) & \frac{1}{3}(\cos\theta-1) & \frac{1}{3}(\cos\theta-1) \\ -\frac{\sin\theta}{\sqrt{3}} & \frac{1}{3}(\cos\theta-1) & \frac{1}{3}(\cos\theta+2) & \frac{1}{3}(\cos\theta-1) \\ -\frac{\sin\theta}{\sqrt{3}} & \frac{1}{3}(\cos\theta-1) & \frac{1}{3}(\cos\theta-1) & \frac{1}{3}(\cos\theta+2) \end{pmatrix}, \quad (1)$$

which represents a 4D rotation by $\theta$ in the plane spanned by $\{1,1,1,1\}$ and $\{1,0,0,0\}$. For the special case of $\theta = \frac{\pi}{3}$, one gets:

$$U_{\text{rot}} = \frac{1}{6}\begin{pmatrix} 3 & 3 & 3 & 3 \\ -3 & 5 & -1 & -1 \\ -3 & -1 & 5 & -1 \\ -3 & -1 & -1 & 5 \end{pmatrix}, \quad (2)$$

Eq.(2) transforms a probability vector $\{p_1, p_2, p_3, 1-(p_1+p_2+p_3)\}$ into $\{\frac{1}{2}, -\frac{1+2p_1}{6}+p_2, -\frac{1+2p_1}{6}+p_3, \frac{1}{6}(5-8p_1-6p_2-6p_3)\}$. With the first element as a constant, one can drop it and reduce the dimension from 4D to 3D. In a similar manner Eq.(2) transforms the original 3-simplex into a new one spanned by $\{\frac{1}{2}, -\frac{1}{2}, -\frac{1}{2}, -\frac{1}{2}\}$, $\{\frac{1}{2}, \frac{5}{6}, -\frac{1}{6}, -\frac{1}{6}\}$, $\{\frac{1}{2}, -\frac{1}{6}, \frac{5}{6}, -\frac{1}{6}\}$, and $\{\frac{1}{2}, -\frac{1}{6}, -\frac{1}{6}, \frac{5}{6}\}$. With all first elements as $\frac{1}{2}$, the transformed simplex can be projected into $\mathbb{R}^3$ space, as a tetrahedron spanned by $\{-\frac{1}{2}, -\frac{1}{2}, -\frac{1}{2}\}$, $\{\frac{5}{6}, -\frac{1}{6}, -\frac{1}{6}\}$, $\{-\frac{1}{6}, \frac{5}{6}, -\frac{1}{6}\}$, and $\{-\frac{1}{6}, -\frac{1}{6}, \frac{5}{6}\}$[1]. We investigate what region of this simplex (i.e. tetrahedron) can be reduced to lower dimensional 1-simplex, implying the original probability 4-vector can be written as product of two disjoint probability vectors in lower dimensions.

Take a 1-simplex spanned by points $\{1, 0\}$ and $\{0, 1\}$ (Figure 1a). Any point sampled from this simplex have the form $\{p, 1-p\}$ with $0 \leq p \leq 1$. Consider two points randomly sampled from this 1-simplex: $\{p, 1-p\}$ and $\{q, 1-q\}$. Their outer product will be given by:

$$\{pq, p(1-q), q(1-p), (1-p)(1-q)\}. \tag{3}$$

Eq.(2) transforms Eq.(3) into:

$$\left\{\frac{1}{2}, -\frac{1}{6} + p - \frac{4pq}{3}, -\frac{1}{6} + q - \frac{4pq}{3}, \frac{5}{6} - p - q + \frac{2pq}{3}\right\}. \tag{4}$$

After dropping the first element, one gets:

$$\left\{-\frac{1}{6} + p - \frac{4pq}{3}, -\frac{1}{6} + q - \frac{4pq}{3}, \frac{5}{6} - p - q + \frac{2pq}{3}\right\} \tag{5}$$

Eq.(5) corresponds to a three-dimensional surface parameterized by variables $0 \leq p \leq 1$ and $0 \leq q \leq 1$. This surface is fully contained within the tetrahedron of probability space, as shown Figure 1b. Not all of the points within the tetrahedron correspond to *physical* quantum states (i.e. states with a positive semi-definite density matrix). The only region of this tetrahedron that corresponds to physical quantum state is the sphere that is inscribed within it[2].

The intersection of the insphere and the surface in Eq.(5), is what we called as *quantum potato chip*. The border of the quantum potato chip, as highlighted in Figure 1b by the solid red line, is described by:

$$q = \frac{1}{2}\left(1 \pm \sqrt{\frac{-1 + 6p - 6p^2}{3(1 - 2p + 2p^2)}}\right). \tag{6}$$

Therefore, the quantum potato chip is a surface described in Eq.(5) and parameterized by $p, q$ as follow:

$$\frac{1}{2}\left(1 - \frac{1}{\sqrt{3}}\right) \leq p \leq \frac{1}{2}\left(1 + \frac{1}{\sqrt{3}}\right), \quad \frac{1}{2}\left(1 - \sqrt{\frac{-1 + 6p - 6p^2}{3(1 - 2p + 2p^2)}}\right) \leq q \leq \frac{1}{2}\left(1 + \sqrt{\frac{-1 + 6p - 6p^2}{3(1 - 2p + 2p^2)}}\right). \tag{7}$$

Additionally, the original 1-simplex in Figure 1b is in fact a 1D object embedded in 3D, spanned by points $\{\{1, 0, 0\}, \{0, 1, 0\}\}$. There are two other choices for points as $\{\{1, 0, 0\}, \{0, 0, 1\}\}$, and $\{\{0, 1, 0\}, \{0, 0, 1\}\}$. Therefore, overall there are three potato chips, as shown in Figure 1c, and their surfaces are described by Eq.(5), Eq.(8) and Eq.(9), with $p, q$ parametrized in Eq.(7).[3]

$$\left\{-\frac{1}{6} + q - \frac{4pq}{3}, \frac{5}{6} - p - q + \frac{2pq}{3}, -\frac{1}{6} + p - \frac{4pq}{3}\right\} \tag{8}$$

---

[1] If one treats this vectors as Bloch vectors, then after normalizing them, one gets a new basis which is equivalent to the Feynman basis (up-to a rotation and a reflection, see Figure 3c for more details). The corresponding orientation is taken to be the default one in this paper.

[2] The insphere, centered at the origin $\{0, 0, 0\}$, has a radius $\frac{1}{2\sqrt{3}}$. This radius is not unity because, for instance, a normalized vector such as $\{1, 0, 0, 0\}$ transforms into the vector $\{\frac{1}{2}, -\frac{1}{2}, -\frac{1}{2}, -\frac{1}{2}\}$, which, after dropping the first component, becomes $\{-\frac{1}{2}, -\frac{1}{2}, -\frac{1}{2}\}$. To maintain normalization, this vector must be rescaled by a factor of $\frac{\sqrt{3}}{2}$. After rescaling, the resulting sphere has radius 1, analogous to the Bloch sphere.

[3] Any two potato chips are related to the third one by two different ways of applying CNOT matrix (corresponding to permutations $\sigma_1 = \begin{pmatrix} 1 & 2 & 3 & 4 \\ 2 & 1 & 3 & 4 \end{pmatrix}$ and $\sigma_2 = \begin{pmatrix} 1 & 2 & 3 & 4 \\ 1 & 2 & 4 & 3 \end{pmatrix}$) to the probability vector.

$$\left\{-\frac{1}{6} - \frac{p}{3} + q - \frac{2pq}{3}, -\frac{1}{6} - \frac{p}{3} + \frac{4pq}{3}, \frac{5}{6} - \frac{4p}{3} - q + \frac{4pq}{3}\right\} \tag{9}$$

## 2.2. Bloch Sphere

Consider the following SIC-POVM for qubits (further referred to as the QBism SIC-POVM)[4], generated from a fiducial vector $\left\{e^{-i\frac{3\pi}{4}}(\sqrt{3}-1), 1\right\}$[5] and POVM elements as follows:

$$Q_1 = \frac{1}{12}\begin{pmatrix} 3-\sqrt{3} & \sqrt{6}e^{-i\frac{3\pi}{4}} \\ \sqrt{6}e^{i\frac{3\pi}{4}} & 3+\sqrt{3} \end{pmatrix}, \quad Q_2 = \frac{1}{12}\begin{pmatrix} 3-\sqrt{3} & \sqrt{6}e^{i\frac{\pi}{4}} \\ \sqrt{6}e^{-i\frac{\pi}{4}} & 3+\sqrt{3} \end{pmatrix},$$

$$Q_3 = \frac{1}{12}\begin{pmatrix} 3+\sqrt{3} & \sqrt{6}e^{i\frac{3\pi}{4}} \\ \sqrt{6}e^{-i\frac{3\pi}{4}} & 3-\sqrt{3} \end{pmatrix}, \quad Q_4 = \frac{1}{12}\begin{pmatrix} 3+\sqrt{3} & \sqrt{6}e^{-i\frac{\pi}{4}} \\ \sqrt{6}e^{i\frac{\pi}{4}} & 3-\sqrt{3} \end{pmatrix}, \tag{10}$$

where $\sum_i Q_i = \mathbb{I}$ and $Q_i \geq 0$. These POVM elements correspond to the following Bloch vectors, respectively[6]: $\left\{-\frac{1}{\sqrt{3}}, \frac{1}{\sqrt{3}}, -\frac{1}{\sqrt{3}}\right\}$, $\left\{\frac{1}{\sqrt{3}}, -\frac{1}{\sqrt{3}}, -\frac{1}{\sqrt{3}}\right\}$, $\left\{-\frac{1}{\sqrt{3}}, -\frac{1}{\sqrt{3}}, \frac{1}{\sqrt{3}}\right\}$, $\left\{\frac{1}{\sqrt{3}}, \frac{1}{\sqrt{3}}, \frac{1}{\sqrt{3}}\right\}$ (see Figure 2a and Figure 2b). Likewise, the corresponding phase-space basis matrix for this SIC-POVM is given by:

$$\mathcal{B} = \left\{ \begin{pmatrix} \frac{1-\sqrt{3}}{2} \\ e^{-i\frac{3\pi}{4}}\sqrt{\frac{3}{2}} \\ e^{i\frac{3\pi}{4}}\sqrt{\frac{3}{2}} \\ \frac{1+\sqrt{3}}{2} \end{pmatrix}, \begin{pmatrix} \frac{1-\sqrt{3}}{2} \\ e^{i\frac{\pi}{4}}\sqrt{\frac{3}{2}} \\ e^{-i\frac{\pi}{4}}\sqrt{\frac{3}{2}} \\ \frac{1+\sqrt{3}}{2} \end{pmatrix}, \begin{pmatrix} \frac{1+\sqrt{3}}{2} \\ e^{i\frac{3\pi}{4}}\sqrt{\frac{3}{2}} \\ e^{-i\frac{3\pi}{4}}\sqrt{\frac{3}{2}} \\ \frac{1-\sqrt{3}}{2} \end{pmatrix}, \begin{pmatrix} \frac{1+\sqrt{3}}{2} \\ e^{-i\frac{\pi}{4}}\sqrt{\frac{3}{2}} \\ e^{i\frac{\pi}{4}}\sqrt{\frac{3}{2}} \\ \frac{1-\sqrt{3}}{2} \end{pmatrix} \right\}. \tag{11}$$

For a probability vector $\vec{p} = \{p_1, p_2, p_3, 1 - (p_1 + p_2 + p_3)\}$ in the basis of Eq.(11), the vectorized density matrix is given by:

$$\vec{\rho} = \mathcal{B}.\vec{p} = \begin{pmatrix} -\sqrt{3}(p_1 + p_2) + \frac{1+\sqrt{3}}{2} \\ \sqrt{\frac{3}{2}}e^{i\frac{3\pi}{4}}((1+i)p_1 + (1-i)p_2 + 2p_3 - 1) \\ \sqrt{\frac{3}{2}}e^{-i\frac{3\pi}{4}}((1-i)p_1 + (1+i)p_2 + 2p_3 - 1) \\ \sqrt{3}(p_1 + p_2) + \frac{1-\sqrt{3}}{2} \end{pmatrix}, \tag{12}$$

whose eigenvalues are:

$$\frac{1}{2}\left(1 \pm \sqrt{3}\sqrt{8(p_1^2 + p_1(p_2 + p_3 - 1) + p_2^2 + p_2 p_3 + p_3^2) - 8p_2 - 8p_3 + 3}\right). \tag{13}$$

By comparing the vectorized density matrix in the Bloch sphere, $\frac{1}{2}\{1+z, x-iy, x+iy, 1-z\}$, with its counterpart in the phase-space as expressed in Eq.(12), one finds that the probability vector $\vec{p}$ is related to the Bloch vector $\{x, y, z\}$ by the following replacement rule and its inverse:

$$\begin{pmatrix} p_1 \\ p_2 \\ p_3 \end{pmatrix} \to \frac{1}{4}\begin{pmatrix} 1 - \frac{1}{\sqrt{3}}(x - y + z) \\ 1 + \frac{1}{\sqrt{3}}(x - y - z) \\ 1 - \frac{1}{\sqrt{3}}(x + y - z) \end{pmatrix}. \tag{14} \qquad \begin{pmatrix} x \\ y \\ z \end{pmatrix} \to \sqrt{3}\begin{pmatrix} 1 - 2p_1 - 2p_3 \\ 1 - 2p_2 - 2p_3 \\ 1 - 2p_1 - 2p_2 \end{pmatrix}. \tag{15}$$

Imposing positivity on the eigenvalues in Eq.(13) and applying the transformation in Eq.(14) yields the condition $\sqrt{x^2 + y^2 + z^2} \leq 1$ which describes the insphere as the only region within the tetrahedron that corresponds to physical quantum states.

---

[4] Any other SIC-POVM for qubits is simply a rotational transformation of this QBism SIC-POVM.
[5] By applying Weyl-Heisenberg displacement operators $X^p Z^q$ for $p, q \in \{0, 1\}$.
[6] More precisely its projectors: $\Pi_i = 2Q_i$. All Bloch vectors are computed from normalized density matrices ($\text{tr}(\rho) = 1$).

The previous geometric description of states in the quantum potato chip region can be applied directly to the Bloch sphere. Define a qubit state by the probability vector in Eq.(3) in the QBism SIC-POVM. Find the state in the Hilbert space using Weyl–Wigner transformation. Set one of the eigenvalues of the density matrix as zero and solve for $q$. The result will be the same as Eq.(6). Replacing this condition into the Bloch vector $\vec{r}$, one gets:

$$\vec{r} = \left\{ -\sqrt{\frac{2}{1+2p(p-1)} - 3}, (\mp 1 \pm 2p)\sqrt{\frac{2}{1+2p(p-1)} - 3}, (2p-1)\sqrt{3} \right\}. \quad (16)$$

As shown in Figure 2b, Eq.(16) parametrizes the boundary of the quantum potato chip in the Bloch sphere (with $0 \leq p \leq 1$), similar to Figure 1b.

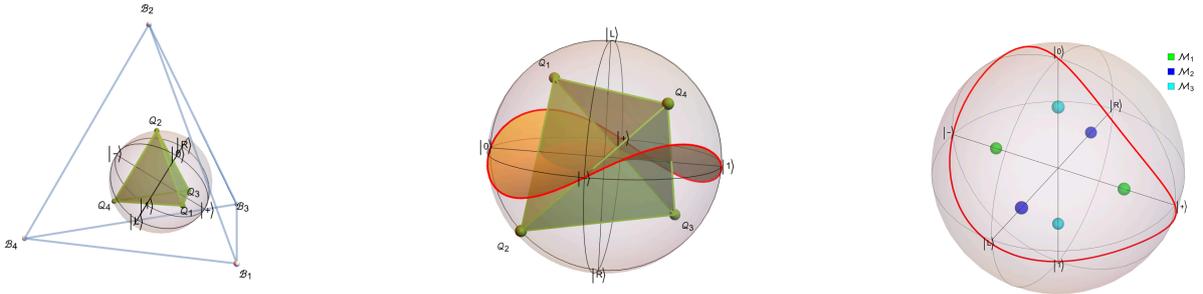

(a) QBism SIC-POVM basis element and its POVM elements in Bloch sphere.

(b) Quantum potato chip and QBism SIC-POVM elements in Bloch sphere.

(c) Quantum potato chips and three new POVM elements.

Figure 2: (a) SIC-POVM basis ($\mathcal{B}_i$, blue) vs POVM elements ($\mathcal{Q}_i$, green), represented in the Bloch sphere. Note the $\mathcal{B}_i$ tetrahedron represents the available phase-space, spanned by QBism SIC-POVM basis, while the POVM elements are within the insphere. (b) quantum potato chip in the Bloch sphere. The green spheres are Bloch vectors for the QBism SIC-POVM in Eq.(11). (c) the solid red line is the boundary of quantum potato chip, while sphere represents new POVM defined in Eq.(17).

## 3. Important features and properties of the quantum potato chip

We investigate the properties of quantum states within the "quantum potato chip", which are free of correlation between its complete projective measurements. We show how states lying on the quantum potato chip can be reconstructed solely through projective measurements (for example measurements in the Pauli-X and Pauli-Z bases). Additionally, we demonstrate that the Matthews correlation measure vanishes exclusively for the quantum potato chip states. We investigate how some common noise channels preserve the quantum potato chip. We also study how a Liouvillian dynamics can keep a state within the quantum potato chip.

### 3.1. Construction of Quantum State by Projective Measurements

#### 3.1.1. SIC-POVM case

A qubit state can be reconstructed from SIC-POVM probabilities, because they are informationally complete (IC)[1–6]. However, general qubit states cannot be recovered from only two independent projective measurements. We will show that for states within the quantum potato chips, it is possible to reconstruct the entire state from only two independent projective measurements. This is related to the fact that their probability space can be reduced into two disjoint separable ones.

Given the QBism SIC-POVM such as Eq.(10), define three new POVM sets as follows: $\mathcal{M}_x$: $\{Q_1 + Q_3, \mathbb{I} - (Q_1 + Q_3)\}$, $\mathcal{M}_y$: $\{Q_1 + Q_4, \mathbb{I} - (Q_1 + Q_4)\}$, and $\mathcal{M}_z$: $\{Q_1 + Q_2, \mathbb{I} - (Q_1 + Q_2)\}$. They correspond to measurements along these Bloch vectors:

$$\mathcal{M}_x : \left\{ \begin{pmatrix} \frac{1}{\sqrt{3}} \\ 0 \\ 0 \end{pmatrix}, \begin{pmatrix} -\frac{1}{\sqrt{3}} \\ 0 \\ 0 \end{pmatrix} \right\}, \mathcal{M}_y : \left\{ \begin{pmatrix} 0 \\ \frac{1}{\sqrt{3}} \\ 0 \end{pmatrix}, \begin{pmatrix} 0 \\ -\frac{1}{\sqrt{3}} \\ 0 \end{pmatrix} \right\}, \mathcal{M}_z : \left\{ \begin{pmatrix} 0 \\ 0 \\ \frac{1}{\sqrt{3}} \end{pmatrix}, \begin{pmatrix} 0 \\ 0 \\ -\frac{1}{\sqrt{3}} \end{pmatrix} \right\}, \qquad (17)$$

This feature is also visualized in Figure 2c. One can see the vectors corresponding to $\mathcal{M}_i$ are along the Cartesian axes (i.e. the same direction for Pauli matrices), but scaled by a factor of $1/\sqrt{3}$.

For a qubit state defined by the probability vector of Eq.(3) in the SIC-POVM basis of Eq.(11), the measurement probabilities for $\mathcal{M}_x$ is $\{q, 1-q\}$ and for $\mathcal{M}_z$ is $\{p, 1-p\}$. Additionally, $\mathcal{M}_x$, $\mathcal{M}_y$ and $\mathcal{M}_z$ can be treated as scaled versions of projective measurements of Pauli operators $\sigma_x$, $\sigma_y$ and $\sigma_z$, respectively. For a generic state such as $\frac{1}{2}(\mathbb{I} + \vec{r}.\vec{\sigma})$, the Pauli-Z probabilities are $\{\frac{1}{2}(1-z), \frac{1}{2}(1+z)\}$, while for $\mathcal{M}_z$ the probabilities are $\{\frac{1}{6}(3-\sqrt{3}z), \frac{1}{6}(3+\sqrt{3}z)\}$; likewise for Pauli-X, one gets $\{\frac{1}{2}(1-x), \frac{1}{2}(1+x)\}$ while those of $\mathcal{M}_x$ are $\{\frac{1}{6}(3-\sqrt{3}x), \frac{1}{6}(3+\sqrt{3}x)\}$. These results can be also found directly from the density matrix in the QBism SIC-POVM basis:

$$\frac{1}{4\sqrt{3}} \begin{pmatrix} (\sqrt{3} - x + y - z) & (\sqrt{3} + x - y - z) \\ (\sqrt{3} - x - y + z) & (\sqrt{3} + x + y + z) \end{pmatrix}. \qquad (18)$$

Adding up columns in Eq.(18) return $\mathcal{M}_z$ probabilities, while for rows summation one gets $\mathcal{M}_x$ probabilities. Given any projective measurement for Pauli matrices, if its probability is denoted by $P(\sigma_i) = p$, the corresponding $\mathcal{M}_i$ probability will be given by $P(\mathcal{M}_i) = \frac{1}{\sqrt{3}}\left(p - \frac{1}{2}\right) + \frac{1}{2}$. This scaling can also be expressed with the following doubly-stochastic matrix[7]:

$$\mathcal{S} = \frac{1}{\sqrt{3}}\mathbb{I} + \left(1 - \frac{1}{\sqrt{3}}\right)\frac{1}{2} = \frac{1}{6}\begin{pmatrix} 3 + \sqrt{3} & 3 - \sqrt{3} \\ 3 - \sqrt{3} & 3 + \sqrt{3} \end{pmatrix}. \qquad (19)$$

such that

$$p \rightarrow \mathcal{S}p. \qquad (20)$$

Therefore, for any state within the quantum potato chips, one can perform Pauli-X and Pauli-Z[8] projective measurements and record corresponding probabilities. By transforming these probabilities using Eq.(20), one recovers $p$ and $q$, reconstructing the probability vector in Eq.(3) and fully reconstructing the quantum state. One should notice that this process is doable for states only in the quantum potato chip and not for general states.

Consider one specific example. Given the state vector from Eq.(3) with parameters $p = \frac{1}{3}$ and $q = \frac{2}{5}$, the QBism SIC-POVM probabilities are $\{\frac{2}{15}, \frac{1}{5}, \frac{4}{15}, \frac{2}{5}\}$. For Pauli-Z and Pauli-X projective measurements, the probabilities are $\{\frac{1}{6}(3-\sqrt{3}), \frac{1}{6}(3+\sqrt{3})\}$ and $\{\frac{1}{10}(5-\sqrt{3}), \frac{1}{10}(5+\sqrt{3})\}$, respectively. For measurements $\mathcal{M}_z$ and $\mathcal{M}_x$, or after applying the transformation from Eq.(20), the probabilities are $\{\frac{1}{3}, \frac{2}{3}\}$ and $\{\frac{2}{5}, \frac{3}{5}\}$, respectively. Their outer product also yields $\{\frac{2}{15}, \frac{1}{5}, \frac{4}{15}, \frac{2}{5}\}$.

This factorization allows us to decouple the probability 4-vector in a 3-simplex into two lower-dimensional disjoint distributions in 1-simplices. Consequently, the probability space for a state in the quantum potato chip can be viewed as the product of two independent subspaces, each governed by their respective probability distributions. In practical terms, the system exhibits a decoupling of correlations between the disjoint subspaces, which simplifies the analysis and allows us to treat each subspace independently. Potential implications will be discussed in our future work.

---

[7]https://mathworld.wolfram.com/DoublyStochasticMatrix.html
[8]Alternatively Pauli-X and Pauli-Y or Pauli-Y and Pauli-Z for a differently oriented potato chip.

### 3.1.2. Quasi-Probability Bases

Historically, the phase-space representation of quantum states, such those in bases of Wootters [7] and Feynman [8][9], involves quasi-probability distributions. These representations allow for the treatment of quantum states in a manner similar to classical statistical distributions, but with key distinctions, such as the possibility of negative values. For example, Wootters basis is given by:

$$\mathcal{W} = \left\{ \begin{pmatrix} 1 \\ \frac{e^{-i\frac{\pi}{4}}}{\sqrt{2}} \\ \frac{e^{i\frac{\pi}{4}}}{\sqrt{2}} \\ 0 \end{pmatrix}, \begin{pmatrix} 0 \\ \frac{e^{i\frac{\pi}{4}}}{\sqrt{2}} \\ \frac{e^{-i\frac{\pi}{4}}}{\sqrt{2}} \\ 1 \end{pmatrix}, \begin{pmatrix} 1 \\ -\frac{e^{-i\frac{\pi}{4}}}{\sqrt{2}} \\ -\frac{e^{i\frac{\pi}{4}}}{\sqrt{2}} \\ 0 \end{pmatrix}, \begin{pmatrix} 0 \\ \frac{e^{i\frac{\pi}{4}}}{\sqrt{2}} \\ -\frac{e^{-i\frac{\pi}{4}}}{\sqrt{2}} \\ 1 \end{pmatrix} \right\}. \quad (21)$$

In Wootters basis, the density matrix for a Bloch vector $\{x, y, z\}$ is described by:

$$\begin{pmatrix} \frac{1}{4}(1+x+y+z) & \frac{1}{4}(1+x-y-z) \\ \frac{1}{4}(1-x-y+z) & \frac{1}{4}(1-x+y-z) \end{pmatrix}. \quad (22)$$

The summation of elements in Eq.(22) along columns results in $\frac{1}{2}(1+z)$ and $\frac{1}{2}(1-z)$, which represent the probabilities associated with projective measurements along the Pauli-Z axis. Similarly, summing the elements along rows yields $\frac{1}{2}(1+x)$ and $\frac{1}{2}(1-x)$, which correspond to the probabilities of measurements along the Pauli-X axis[10]. In other words, in this case, there is no need for scaling or applying any extra transformation as in Eq.(20) for SIC-POVM case.

Wootters phase-space basis is no longer a proper probability distribution, and it is usually referred to as *quasi*-distribution. This provides another description of quantum potato chips as states that correspond to the subset of quasi-distributions with all entries being positive.

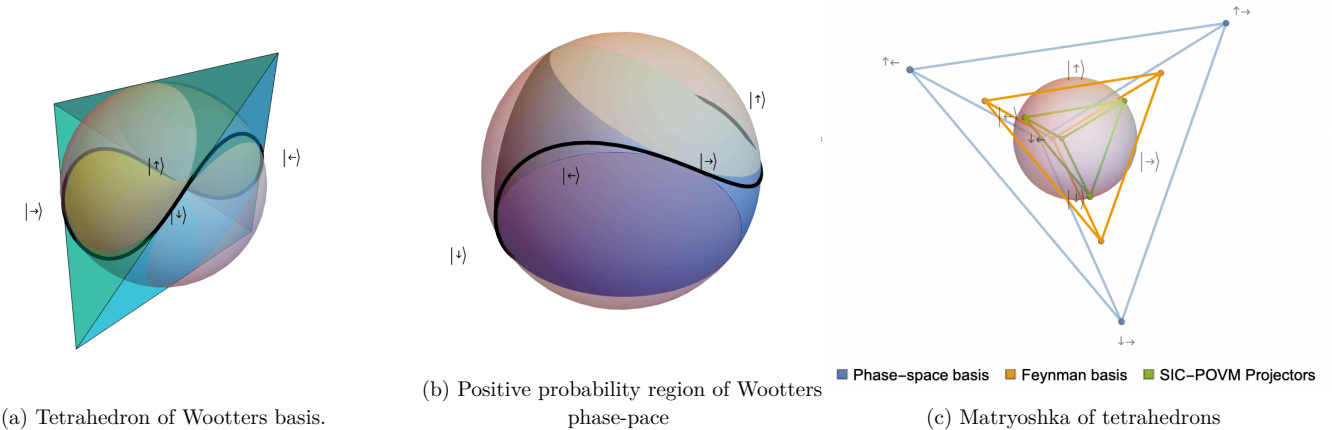

(a) Tetrahedron of Wootters basis.   (b) Positive probability region of Wootters phase-pace   (c) Matryoshka of tetrahedrons

Figure 3: (a) the tetrahedron represents points in the phase-space of Wootters basis that corresponds to only positive probabilities. Note the Bloch sphere is not inscribed within the tetrahedron and not all points within the tetrahedron represents quantum states. Pauli-Z eigenstates are $|\uparrow\rangle, |\downarrow\rangle$ and X are $|\leftarrow\rangle, |\rightarrow\rangle$. (b) Quantum potato chip represents the states with independent observables, but represented in the Wootters basis. Its boundary is the solid black line. The region inside the Bloch sphere is the intersection of the Wootters tetrahedron and Bloch sphere, corresponding to quantum states represented by positive probability vectors[11] (c) The correspondence between the probabilistic phase-space basis (blue), quasi-probabilistic Feynman basis [8] (orange) and SIC-POVM projectors. Each tetrahedron is a scaled down version of the previous one by a factor of $\sqrt{3}$ (green). Each corner is denoted by two arrows, representing spins along z and x axes (up/down for z-axis and right/left for x-axis).

Allowing quasi probabilities has some geometric consequences. If one considers only positive (say, classical) probabilities, the corresponding tetrahedron does not completely contain the Bloch sphere (see

---

[9]For qubits Feynman and Wootters bases are the same and we refer to both interchangeably.

[10]Pauli-Z and Pauli-X measurement can be interpreted as the position (Pauli-Z) and momentum (Pauli-X) measurements within the phase space.

[11]Refer to Fig.4 in [10], which has a similar visualization, in a slightly different context.

Figure 3). In other words, parts of Bloch sphere outside of the tetrahedron correspond to states with negative probabilities in this basis. Additionally, the parametrization of the quantum potato chip will be different compared to Eq.(6):

$$q = \frac{1}{2} \pm \sqrt{\frac{p(1-p)}{2(1-2p+2p^2)}}. \tag{23}$$

A similar factorization argument for the probability vectors of the quantum potato chip can be applied in this context. However, it is crucial to note that, unlike the SIC-POVM case, the probabilities here do not necessarily correspond to directly measurable quantities, as they may take on negative values.

### 3.2. Matthews correlation of classical binary variables

To quantify the correlation of classical binary variables within the quantum potato chip compared to other points in the Bloch sphere, one can use the Matthews correlation coefficient. Given binary variables defining the probability vector as:

$$\vec{p} = \{p_1, p_2, p_3, p_4\} \to \begin{pmatrix} p_{11} & p_{12} \\ p_{21} & p_{22} \end{pmatrix}, \tag{24}$$

the corresponding Matthews correlation coefficient is given by:

$$\varphi = \frac{p_{22}p_{11} - p_{12}p_{21}}{\sqrt{(p_{11}+p_{12})(p_{21}+p_{22})(p_{11}+p_{21})(p_{12}+p_{22})}} = \frac{\sqrt{3}y - xz}{\sqrt{(3-x^2)(3-z^2)}}, \tag{25}$$

where in the last step, we replace variables by the Bloch vector components $\{x, y, z\}$ using Eq.(15).

As shown in Figure 4, the only region with $\varphi = 0$ is the quantum potato chip. This means that measurement predictions in Eq.(24), which can be obtained experimentally by measuring POVMs, are uncorrelated, much like flipping a pair of independent biased coins (probabilities are split as described by Eq.(3)). Equivalently, the states within the quantum potato chips can be represented as a product state in terms of the measurement basis associated with a SIC-POVM; precisely the scenario where the qubit state does not exhibit quantum correlations between certain SIC-POVM measurement outcomes.

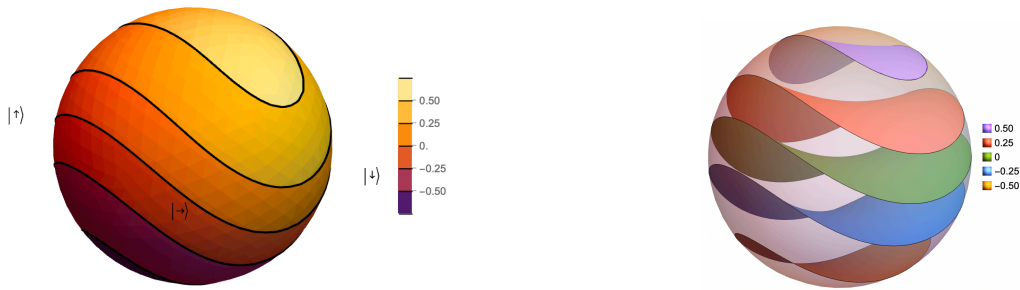

Figure 4: Matthews correlation coefficient Eq.(25) for the pair of binary variables forming the probability vector Eq.(24). $\varphi$ has minimum $-\frac{1}{\sqrt{3}}$ and maximum $\frac{1}{\sqrt{3}}$ at the poles of the sphere

The Eq.(25) takes a more natural form if Bloch sphere's radius[12] is $\frac{1}{\sqrt{3}}$:

$$\varphi = \frac{y - xz}{\sqrt{(1-x^2)(1-z^2)}}, \tag{26}$$

---

[12]Read our footnote for insphere, after Eq.(5)

## 3.3. Effect of Error Channels on the Quantum Potato Chip

In this section, we explore the impact of noise on the quantum potato chip. Specifically, we model various types of errors using known quantum channels. See Table 1 for detailed description of quantum channels, their Kraus operators and how they transform the Bloch sphere.

| Channel name | List of Kraus operators | Channel's effect on Bloch sphere |
|---|---|---|
| Bit flip | $\{\sqrt{\xi}\,\sigma_x, \sqrt{1-\xi}\,\mathbb{I}\}$ | |
| Phase flip | $\{\sqrt{\xi}\,\sigma_z, \sqrt{1-\xi}\,\mathbb{I}\}$ | |
| Bit-phase flip | $\{\sqrt{\xi}\,\sigma_y, \sqrt{1-\xi}\,\mathbb{I}\}$ | |
| Depolarization | $\{\sqrt{\frac{\xi}{4}}\,\sigma_x, \sqrt{\frac{\xi}{4}}\,\sigma_y, \sqrt{\frac{\xi}{4}}\,\sigma_z, \sqrt{1-\frac{3\xi}{4}}\,\mathbb{I}\}$ | |
| Amplitude damping | $\{\frac{1-\sqrt{\xi}}{2}\sigma_z + \frac{1+\sqrt{1-\xi}}{2}\mathbb{I}, \frac{\sqrt{\xi}}{2}\sigma_x + i\frac{\sqrt{\xi}}{2}\sigma_y\}$ | |
| Phase damping | $\{\frac{1-\sqrt{1-\xi}}{2}\sigma_z + \frac{1+\sqrt{1-\xi}}{2}\mathbb{I}, \frac{\sqrt{\xi}}{2}\mathbb{I} - \frac{\sqrt{\xi}}{2}\sigma_z\}$ | |

Table 1: Common noise channels and their corresponding Kraus operators. We also show how each one changes Bloch sphere.

The Bloch surface of the quantum potato chip, $\sqrt{3}\{(1-2q), (2p-1)(2q-1), (1-2p)\}$ will be transformed into a new one as shown in the following equation, for bit flip, phase flip, bit-phase flip, depolarization, amplitude damping, and phase damping, respectively:

| Channel name | Resultant Bloch vector | Channel's effect on Potato chip |
|---|---|---|
| Bit flip | $\sqrt{3}\{-f_q, -f_p f_q f_\xi, f_p f_\xi\}$ | |
| Phase flip | $\sqrt{3}\{f_q f_\xi, -f_p f_q f_\xi, -f_p\}$ | |
| Bit-phase flip | $\sqrt{3}\{-f_q f_\xi, f_p f_q, f_p f_\xi\}$ | |
| Depolarization | $\sqrt{3}\{-f_q(1-\xi), f_p f_q(1-\xi), -f_p(1-\xi)\}$ | |
| Amplitude damping | $\sqrt{3}\{-f_q(1-\xi), f_p f_q \sqrt{1-\xi}, \frac{\xi}{\sqrt{3}} - f_p(1-\xi)\}$ | |
| Phase damping | $\sqrt{3}\{-f_q \sqrt{1-\xi}, f_p f_q \sqrt{1-\xi}, -f_p\}$ | |

Table 2: Channels' effects on potato chip region. Considering $f_p = 2p-1$, $f_q = 2q-1$, $f_\xi = 2\xi-1$. Image: We set error rate/probability as $\xi = \frac{1}{3}$. The only noise channels that keep states within the quantum potato chips are bit flip, phase flip and phase damping.

Upon a close analysis, one can show that bit flip, phase flip and phase damping map the states within the quantum potato chip to other states within the quantum potato chip, only reducing the volume of the region (see Table 2). On the other hand, other channels map states such that they go outside of the quantum potato chip. To explicitly show that the probabilities are still separable for the bit flip case, the Bloch vector can be obtained by new variables $q', p'$ and Eq.(3) with $\{p' = p + \xi(1-2p), q' = q\}$, for the phase flip it will be $\{q' = q + \xi(1-2q), p' = p\}$, while for the phase damping it will be $\{p' = p, q' = \frac{1}{2}(1 - \sqrt{1-\xi} + 2q\sqrt{1-\xi})\}$.

## 3.4. Unconventional Liouvillian Evolution at the Boundary of Quantum Potato Chips

A natural physics question arises: can the boundary of the quantum potato chip be understood as the result of state evolution under a master equation, starting from an appropriate initial condition?

Given that the trajectory of each individual probability vector $P = \{p, 1-p\}$ and $Q = \{q, 1-q\}$ is already known and constrained by Eq.(23), the parameter $p$ can be interpreted as a time variable in a parametric equation for the whole system. We can identify two time (or parameter)-dependent transition matrices that generate the desired trajectory.

$$P'(p) = \mathcal{L}_1(p)P(p), \quad Q'(p) = \mathcal{L}_2(p)Q(p) \tag{27}$$

By combining these two time-dependent transition matrices, we obtain the resulting transition matrix, which governs the overall evolution of the system.

$$\mathcal{L} = \log(e^{\mathcal{L}_1} \otimes e^{\mathcal{L}_2}) \tag{28}$$

Any logarithm of a $2 \times 2$ doubly stochastic matrix depends only on a single parameter, as the requirement that both its rows and columns sum to 0 imposes strict constraints on its form.

$$\mathcal{L}_1 = \begin{pmatrix} -x & x \\ x & -x \end{pmatrix}, \quad \mathcal{L}_2 = \begin{pmatrix} -y & y \\ y & -y \end{pmatrix} \tag{29}$$

We can then simply solve for $x$ and $y$ as follows:

$$x = \frac{1}{1-2p}, \quad y = -\frac{1-2p}{4p(1-p)((1-p)^2 + p^2)}. \tag{30}$$

For $0 < p < \frac{1}{2}$, the term $\exp(\mathcal{L}_1)$ is stochastic and $\exp(\mathcal{L}_2)$ backward-stochastic[13], while for $\frac{1}{2} < p < 1$, the term $\exp(\mathcal{L}_2)$ is stochastic and $\exp(\mathcal{L}_1)$ is backward-stochastic. Opposite direction of inference is what keeps the overall distribution spread (and purity with von Neumann entropy) constant. The Shannon entropy, $H$, of $P \times Q$ would be bounded:

$$1 \leq H \leq 1.35226 < 2 \tag{31}$$

Accordingly, for the overall transition matrix one finds:

$$\mathcal{L} = \begin{pmatrix} \frac{8(p-1)p((p-1)p+1)+1}{4(p-1)p(2p-1)(2(p-1)p+1)} & \frac{1-2p}{4(p-1)p(2(p-1)p+1)} & \frac{1}{1-2p} & 0 \\ \frac{1-2p}{4(p-1)p(2(p-1)p+1)} & \frac{8(p-1)p((p-1)p+1)+1}{4(p-1)p(2p-1)(2(p-1)p+1)} & 0 & \frac{1}{1-2p} \\ \frac{1}{1-2p} & 0 & \frac{8(p-1)p((p-1)p+1)+1}{4(p-1)p(2p-1)(2(p-1)p+1)} & \frac{1-2p}{4(p-1)p(2(p-1)p+1)} \\ 0 & \frac{1}{1-2p} & \frac{1-2p}{4(p-1)p(2(p-1)p+1)} & \frac{8(p-1)p((p-1)p+1)+1}{4(p-1)p(2p-1)(2(p-1)p+1)} \end{pmatrix} \tag{32}$$

In the Hilbert space, the master equation corresponding to the above transition can be expressed as:

$$\dot{\rho} = \gamma_1 \left( L_1 \rho L_1^\dagger - \frac{1}{2}\{L_1^\dagger L_1, \rho\} \right) + \gamma_2 \left( L_2 \rho L_2^\dagger - \frac{1}{2}\{L_2^\dagger L_2, \rho\} \right), \tag{33}$$

with two Lindblad jump operators:

$$L_1 = \frac{1}{\sqrt{1-2p}}(\mathbb{I} + \sigma_z), \quad L_2 = \frac{1}{2}\sqrt{\frac{1-2p}{p(1-p)(1-2p(1-p))}} \, \sigma_x. \tag{34}$$

---

[13] If $\exp(\mathcal{A})$ is stochastic then $\exp(-\mathcal{A})$ is called backward-stochastic.

However, here we have unconventional (negative) opposite damping rates $\gamma_1 = -\gamma_2 = 1$. Additionally, because of the singularity at $p = \frac{1}{2}$, damping rates should swap signs to close the trajectory:

$$\gamma_1 = -\gamma_2 = \begin{cases} 1 \text{ if } 0 < p \leq \frac{1}{2} \\ -1 \text{ if } \frac{1}{2} < p < 1 \end{cases} \quad (35)$$

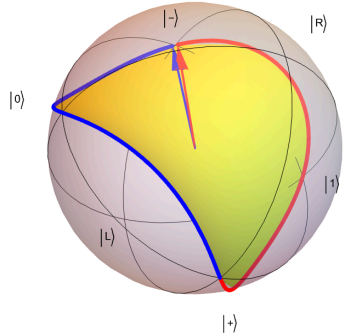

Figure 5: Two solutions from Eq.(23) define opposite shifts to the initial state $|-\rangle \pm \delta p$, evolving in opposite directions.

While this specific evolution may lack an immediate physical interpretation, it is significant that, starting from any point on the boundary of the quantum chip, there is a non-conventional dynamical equation, guaranteeing that the quantum state remains pure physical state confined to it.

## 4. Concluding remarks

SIC-POVMs offer an elegant framework for representing qubit states geometrically, using a 4-dimensional probability vector within a 3-simplex. Through geometric transformations, this simplex can be projected into a regular tetrahedron in $\mathbb{R}^3$. Crucially, not all points in this tetrahedron represent valid qubits, with physical quantum states residing only within the insphere of the tetrahedron. A specific surface within this space, defined by the product of two 1-simplices, representing uncorrelated pair of binary variables, intersects the insphere forming the "quantum potato chip," a subset of quantum states of the qubit with uncorrelated projective observables. We explored a few of important features of states within the quantum potato chip. For example, these states can be fully constructed by one two projective measurements, their corresponding Matthew's correlation is zero, and any binary classical problem can be directly mapped into these states.

The quantum potato chip, in particular its complete description by two binary classical variables, suggests an important question: can universal quantum computation still be achieved after restricting qubits to such states? Exploring this question and possible modifications to quantum formalism will be explored in future work. The geometric structure of SIC-POVMs suggests that understanding these spaces could unlock new pathways in quantum computing and quantum state representation.

## Acknowledgement

Computational aspects of this paper (e.g., mathematical derivation of formulas and visualizations) were done in the version 14.0 of the Wolfram Language. We also extensively used the Wolfram quantum framework, which a library of functionalities to perform symbolic and numeric quantum computation in the discrete finite-dimensional vector space. We thank James Wiles from Wolfram Institute, for his valuable comments on an early draft of our work.

**Eq.(1)**

```
rot = RotationMatrix[θ, {{1, 1, 1, 1}, {1, 0, 0, 0}}];
rot // MatrixForm
```

**Eq.(2)**

```
rotation = rot /. θ -> π/3;
rotation // MatrixForm
```

**Eq.(3)**

```
product = KroneckerProduct[{p, 1 - p}, {q, 1 - q}] // Flatten
```

**Eq.(4)**

```
surface4d = rotation . product // ComplexExpand
```

**Eq.(5)**

```
surface = Rest[surface4d]
```

**Eq.(6)**

```
boundary = Normal[Solve[Norm[2 Sqrt[3] surface] == 1, q, Reals]]
```

**Eq.(7)**

```
{0 <= p <= 1, First[#] <= q <= Last[#] &@@Values[Flatten@boundary]}
```

**Eq.(8) & Eq.(9)**

```
products = Permute[product, #] & /@ {{1, 4, 2, 3}, {3, 1, 2, 4}};
surfaces = ComplexExpand[Rest[rotation . #] & /@ products]
```

**Eq.(24) & Eq.(25)**

```
Phi[dist_CategoricalDistribution] :=
 Phi[Information[dist, "ProbabilityArray"]]
Phi[p_?MatrixQ] /; Dimensions[p] == {2, 2} := \!\(TraditionalForm\`
\*FractionBox[\(p[[2, 2]]\ p[[1, 1]] - p[[1, 2]]\ p[[2, 1]]\),
SqrtBox[\(\(((p[[1, 1]] + p[[1, 2]])\)\ \((p[[2, 1]] +
       p[[2, 2]])\)\ \((p[[1, 1]] + p[[2, 1]])\)\ \((p[[1, 2]] +
       p[[2, 2]])\)\)\)]]\)
FullSimplify@
 Phi@ArrayReshape[
   QuantumPhaseSpaceTransform[QuantumState["BlochVector"[{x, y, z}]],
     basis]["AmplitudesList"], {2, 2}]
```

**Eq.(17) & Eq.(18)**

```
basis = With[{basis = QuantumBasis["QBismSIC"]},
  QuantumBasis[basis["Names"],
   Threaded[IdentityMatrix[2]] - basis["Elements"] /.
```

```
      x_?NumericQ :> (#1 + #2 I & @@ RootApproximant[ReIm[x]])]];
povm = QuantumMeasurementOperator[
  Inverse[Outer[Tr@*Dot, #, #, 1]] . # &@basis["Elements"]];
ops = With[{povm = povm["POVMElements"]},
   With[{op = povm[[#1]] + povm[[#2]]},
      Simplify@{Normal[op], IdentityMatrix[2] - op}] & @@@ {{1,
      2}, {1, 3}, {1, 4}}];
Map[(QuantumState[#]["BlochVector"] // Simplify) &, ops, {2}]
densityQBISMSIC =
 ArrayReshape[
  With[{povms = povm["POVMElements"],
    s = QuantumState["BlochVector"[{x, y, z}]][
      "DensityMatrix"]}, (Tr[# . s] // FullSimplify) & /@ povms], {2,
   2}]
```

### Figure 1a

```
Graphics[{                                                            WL
  {Text[Style["{q,1-q}", 10], {.37, .83}],
   Text[Style["{p,1-p}", 10], {.7, .45}]},
  {ColorData[97][1], Simplex[{{1, 0}, {0, 1}}]}, {Black,
   Point[{.25, .75}]},
  {Red, Point[{.65, .35}]}
  },
 Axes -> True, AxesOrigin -> {0, 0},
 AxesLabel -> {"p", "1-p"}, PlotLabel -> "1-simplex",
 Frame -> True, GridLines -> Automatic,
 FrameTicks -> {{Automatic, None}, {Automatic, None}}
 ]
```

### Figure 1b

```
Show[                                                                 WL
  Graphics3D[
   {
    {Opacity[.5], Ball[{0, 0, 0}, 1/(2 Sqrt[3])]},
    {ColorData[97][1], Thickness[.005], Simplex[{{1, 0, 0}, {0, 1, 0}}]}
   },

    Axes -> True,
    AxesOrigin -> {0, 0, 0},
    Boxed -> False
   ],

  ParametricPlot3D[surface /. boundary, {p, 0, 1}, Mesh -> None, PlotStyle -> Red],

  ParametricPlot3D[surface, {p, 0, 1}, {q, 0, 1}, PlotStyle -> {Opacity[0.6]}],
```

```
  ImageSize -> 400]
```

**Figure 1c**

```
ParametricPlot3D[
 Evaluate[Prepend[surfaces, surface]], {p, 0, 1}, Prepend[Flatten[Values[boundary]], q],
  Mesh -> None,
  PlotStyle -> Opacity[.5],
  PlotPoints -> 200,
  Boxed -> False,
  Axes -> False,
  ImageSize -> 200
 ]
```

**Figure 2a**

```
(*blochsphere is given in previous examples*)
basisPnts =
  QuantumState[#]["BlochVector"] & /@
    QuantumBasis["QBismSIC"]["Elements"];
qbism = QuantumState[#]["BlochVector"] & /@
    QuantumMeasurementOperator["QBismSICPOVM"]["POVMElements"] //
   Simplify;
Show[blochsphere,
 Graphics3D[{
    {MapIndexed[{PointSize[0.01], Sphere[#1, .05], Black,
       Text[Subscript["\[ScriptCapitalB]", #2[[1]]], 1.05 #1]} &,
     basisPnts],
    {Opacity[.2], FaceForm[None],
     EdgeForm[{Opacity[.5], Thick, ColorData[97][1]}],
     Tetrahedron[basisPnts]}},
   {
    ColorData[97][3],
    MapIndexed[{PointSize[0.01], Sphere[#1, .05], Black,
       Text[Subscript["\[ScriptCapitalQ]", #2[[1]]], 1.2 #1]} &,
     qbism],
    {Opacity[.5], EdgeForm[{Thick, ColorData[97][3]}],
     Tetrahedron[qbism]}}
   }, Boxed -> False]]
```

**Figure 2b**

```
blochsphere = Show[
   Graphics3D[{
          { Opacity[0.2], Sphere[]}, Black, Thickness[0.001],
     Opacity[1.0],

     Splice @ {Line[{{0, 1, 0}, {0, -1, 0}}],
       Line[{{0, 0, 1}, {0, 0, -1}}], Line[{{1, 0, 0}, {-1, 0, 0}}]},
```

```
            Splice @ {
                    Text[Ket[{0}], {0, 0, 1.05}],
         Text[Ket[{1}], {0, 0, -1.05}],
                    Text[Ket[{"R"}], {0, 1.05, 0}],
         Text[Ket[{"L"}], {0, -1.05, 0}],
                    Text[Ket[{"+"}], {1.05, 0, 0}],
         Text[Ket[{"-"}], {-1.05, 0, 0}]}
      }, Boxed -> False ],
    ParametricPlot3D[
                 {{Cos[t], Sin[t], 0}, {0, Cos[t], Sin[t]}, {Cos[t],
        0, Sin[t]}},
                 {t, 0, 2 Pi},
                 PlotStyle -> ConstantArray[{Black, Thin}, 3]
             ]];
qbism = {{-(1/Sqrt[3]), 1/Sqrt[3], -(1/Sqrt[3])}, {1/Sqrt[
   3], -(1/Sqrt[3]), -(1/Sqrt[3])}, {-(1/Sqrt[3]), -(1/Sqrt[3]), 1/
   Sqrt[3]}, {1/Sqrt[3], 1/Sqrt[3], 1/Sqrt[3]}};
boundary = {{q ->
     1/2 - Sqrt[(-2 + 12 p - 12 p^2)/(1 - 2 p + 2 p^2)]/(
     2 Sqrt[6])}, {q ->
     1/2 + Sqrt[(-2 + 12 p - 12 p^2)/(1 - 2 p + 2 p^2)]/(
     2 Sqrt[6])}};
blochVector = {Sqrt[3] (-1 + 2 q), Sqrt[3] (-1 + p (2 - 4 q) + 2 q),
   Sqrt[3] (-1 + 2 p)};
Show[
 Graphics3D[{
   ColorData[97][3],
   MapIndexed[{PointSize[0.01], Sphere[#1, .05], Black,
      Text[Subscript["\[ScriptCapitalQ]", #2[[1]]], 1.2 #1]} &, qbism],
   {Opacity[.5], EdgeForm[{Thick, ColorData[97][3]}],
    Tetrahedron[qbism]}}, Boxed -> False],
 ParametricPlot3D[blochVector /. boundary, {p, 0, 1},
  PlotStyle -> Red],
 ParametricPlot3D[
  blochVector, {p, 0, 1}, {q, Sequence @@ boundary[[All, -1, -1]]},
  PlotPoints -> 200, Mesh -> None, PlotStyle -> Opacity[.7]],
 blochsphere
 ]
```

### Figure 2c

```
(*get blochsphere from previous examples*)                                WL
vec1 = {-Sqrt[-3 + 2/(1 + 2 (-1 + p) p)], (-1 + 2  p)  Sqrt[-3 + 2/(
     1 + 2 (-1 + p) p)], Sqrt[3]  (-1 + 2  p)};
vec2 = {Sqrt[-3 + 2/(
    1 + 2 (-1 + p) p)], (1 - 2  p)  Sqrt[-3 + 2/(1 + 2 (-1 + p) p)],
   Sqrt[3]  (-1 + 2  p)};
Legended[Show[
```

```
  blochsphere,
  ParametricPlot3D[{vec1, vec2}, {p, 0, 1}, PlotStyle -> Red],
  Graphics3D[{Opacity[.5],
    Green,
    Sphere[#, .05] & /@ {{1/Sqrt[3], 0, 0}, {-(1/Sqrt[3]), 0, 0}},
    Blue,
    Sphere[#, .05] & /@ {{0, -(1/Sqrt[3]), 0}, {0, 1/Sqrt[3], 0}},
    Cyan, Sphere[#, .05] & /@ {{0, 0, 1/Sqrt[3]}, {0, 0, -(1/Sqrt[3])}}
    }]],
 Placed[SwatchLegend[{Green, Blue,
    Cyan}, {"\!\(\*SubscriptBox[\(\[ScriptCapitalM]\), \(1\)]\)",
    "\!\(\*SubscriptBox[\(\[ScriptCapitalM]\), \(2\)]\)",
    "\!\(\*SubscriptBox[\(\[ScriptCapitalM]\), \(3\)]\)"}], {.95, \
.85}]]
```

**Figure 3a**

```
Show[Graphics3D[{{Text[Ket[{"\[UpArrow]"}], {0, 0, 1.3}],
    Text[Ket[{"\[DownArrow]"}], {0, 0, -1.3}],
    Text[Ket[{"\[RightArrow]"}], {1.3, 0, 0}],
    Text[Ket[{"\[LeftArrow]"}], {-1.3, 0, 0}]}, Opacity[.5], Sphere[],
    Cyan, Simplex[{{1, 1, 1}, {1, -1, -1}, {-1, -1, 1}, {-1,
      1, -1}}]}, Boxed -> False],
 ParametricPlot3D[{{-1 +
     2 p, (-1 + 2 p) (-1 +
      2 (1/2 - Sqrt[-(((-1 + p) p)/(2 - 4 p + 4 p^2))])), -1 +
     2 (1/2 - Sqrt[-(((-1 + p) p)/(2 - 4 p + 4 p^2))])}, {-1 +
     2 p, (-1 + 2 p) (-1 +
      2 (1/2 + Sqrt[-(((-1 + p) p)/(2 - 4 p + 4 p^2))])), -1 +
     2 (1/2 + Sqrt[-(((-1 + p) p)/(2 - 4 p + 4 p^2))])}}, {p, 0, 1},
  PlotStyle -> Directive[Thickness[.01], Black]],
 ParametricPlot3D[{-1 + 2 p, (-1 + 2 p) (-1 + 2 q), -1 +
    2 q}, {p, 0, 1}, {q,
   1/2 - Sqrt[-(((-1 + p) p)/(2 - 4 p + 4 p^2))],
   1/2 + Sqrt[-(((-1 + p) p)/(2 - 4 p + 4 p^2))]}, Mesh -> None,
  PlotStyle -> Yellow]]
```

**Figure 3b**

```
Show[Graphics3D[{{Text[Ket[{"\[UpArrow]"}], {0, 0, 1.3}],
    Text[Ket[{"\[DownArrow]"}], {0, 0, -1.3}],
    Text[Ket[{"\[RightArrow]"}], {1.3, 0, 0}],
    Text[Ket[{"\[LeftArrow]"}], {-1.3, 0, 0}]}, Opacity[.5],
   Sphere[]}, Boxed -> False],
 ParametricPlot3D[{{-1 +
     2 p, (-1 + 2 p) (-1 +
      2 (1/2 - Sqrt[-(((-1 + p) p)/(2 - 4 p + 4 p^2))])), -1 +
     2 (1/2 - Sqrt[-(((-1 + p) p)/(2 - 4 p + 4 p^2))])}, {-1 +
```

```
      2  p, (-1 + 2  p)  (-1 +
        2  (1/2 + Sqrt[-(((-1 + p) p)/(2 - 4 p + 4 p^2))])), -1 +
        2  (1/2 + Sqrt[-(((-1 + p) p)/(2 - 4 p + 4 p^2))])]}, {p, 0, 1},
  PlotStyle -> Directive[Thickness[.01], Black]],
 BoundaryDiscretizeRegion[
  CSGRegion[
   "Intersection", {Ball[],
    Simplex[{{1, 1, 1}, {1, -1, -1}, {-1, -1, 1}, {-1, 1, -1}}]}],
  PrecisionGoal -> 4]]
```

### Figure 3c

```
densityMatrix[phi_] := KroneckerProduct[phi, Conjugate[phi]]
blochVector[rho_?MatrixQ] := Tr /@ (PauliMatrix[Range[3]] . rho)
blochVector[phi_?VectorQ] := blochVector[densityMatrix[phi]]

basis = {{{1/2 (1 - Sqrt[3]), (-(1/2) - I/2) Sqrt[
      3]}, {(-(1/2) + I/2) Sqrt[3],
     1/2 (1 + Sqrt[3])}}, {{1/2 (1 - Sqrt[3]), (1/2 + I/2) Sqrt[
      3]}, {(1/2 - I/2) Sqrt[3],
     1/2 (1 + Sqrt[3])}}, {{1/2 (1 + Sqrt[3]), (-(1/2) + I/2) Sqrt[
      3]}, {(-(1/2) - I/2) Sqrt[3],
     1/2 (1 - Sqrt[3])}}, {{1/2 (1 + Sqrt[3]), (1/2 - I/2) Sqrt[
      3]}, {(1/2 + I/2) Sqrt[3], 1/2 (1 - Sqrt[3])}}};
basisVectors = blochVector /@ basis;
povmProjectors = 2 Inverse[Outer[Tr@*Dot, #, #, 1]] . # &@basis;
feynmanVectors = {{-1, 1, -1}, {1, -1, -1}, {-1, -1, 1}, {1, 1, 1}};
povmVectors = blochVector /@ povmProjectors // Simplify;
labels = {"\[DownArrow]\[LeftArrow]", "\[DownArrow]\[Rule]",
   "\[UpArrow]\[LeftArrow]", "\[UpArrow]\[Rule]"};
Legended[Graphics3D[{
   Opacity[.5],
   Sphere[],
   {{Text[Ket[{"\[UpArrow]"}], {0, 0, 1.3}],
     Text[Ket[{"\[DownArrow]"}], {0, 0, -1.3}],
     Text[Ket[{"\[RightArrow]"}], {1.3, 0, 0}],
     Text[Ket[{"\[LeftArrow]"}], {-1.3, 0, 0}]}},
   {
    ColorData[97][1],
    MapIndexed[{PointSize[0.01], Sphere[#1, .05], Black,
       Text[Subscript["\[ScriptCapitalB]", labels[[#2[[1]]]]],
        1.1 #1]} &, basisVectors], {FaceForm[None],
     EdgeForm[{Opacity[.5], Thick, ColorData[97][1]}],
     Tetrahedron[basisVectors]}
    },
   {
    ColorData[97][2],
    MapIndexed[{PointSize[0.01], Sphere[#1, .05], Black,
```

```
      Text[Subscript["\[ScriptCapitalF]", labels[[#2[[1]]]]],
        1.2 #1]} &, feynmanVectors], {FaceForm[None],
     EdgeForm[{Thick, ColorData[97][2]}],
     Tetrahedron[feynmanVectors]}
    },
    {
     ColorData[97][3],
     MapIndexed[{PointSize[0.01], Sphere[#1, .05], Black,
        Text[Subscript["\[ScriptCapitalQ]", labels[[#2[[1]]]]],
         1.2 #1]} &, povmVectors], {FaceForm[None],
     EdgeForm[{Thick, ColorData[97][3]}], Tetrahedron[povmVectors]}
    }
   }, Boxed -> False],
 SwatchLegend[
  ColorData[97] /@ Range[3], {"Basis", "Feynman", "SIC-POVM"}]
 ]
```

## Figure 4a, Figure 4b

```
Show[
 Graphics3D[{{Text[Ket[{"\[UpArrow]"}], {0, 0, 1.3}],
    Text[Ket[{"\[DownArrow]"}], {0, 0, -1.3}],
    Text[Ket[{"\[RightArrow]"}], {1.3, 0, 0}](*,Text[
    Ket[{"\[LeftArrow]"}],{-1.3,0,0}]*)}}],
 SliceContourPlot3D[(Sqrt[3] y - x z)/Sqrt[(x^2 - 3) (z^2 - 3)],
  "CenterSphere", {x, y, z} \[Element] Ball[], Contours -> 5,
  ColorFunction ->
   Function[x, Opacity[1, ColorData["SunsetColors"][x]]],
  ContourStyle -> Directive[Thick, Black], PlotLegends -> Automatic,
  PlotPoints -> 10],
 Boxed -> False
 ]

ContourPlot3D[(
 Sqrt[3] y - x z)/Sqrt[(x^2 - 3) (z^2 - 3)], {x, -1, 1}, {y, -1,
  1}, {z, -1, 1}, Contours -> Range[1/2, -1/2, -1/4],
 RegionFunction -> Function[{x, y, z}, x^2 + y^2 + z^2 < 1],
 Mesh -> None, RegionBoundaryStyle -> None, Boxed -> False,
 Axes -> False, PlotLegends -> Automatic, PlotRangePadding -> None
 ]
```

## Figure 3a

```
Show[Graphics3D[{{Text[Ket[{"\[UpArrow]"}], {0, 0, 1.3}],
    Text[Ket[{"\[DownArrow]"}], {0, 0, -1.3}],
    Text[Ket[{"\[RightArrow]"}], {1.3, 0, 0}],
    Text[Ket[{"\[LeftArrow]"}], {-1.3, 0, 0}]}, Opacity[.5], Sphere[],
    Cyan, Simplex[{{-((1 + Sqrt[3])/Sqrt[2 (2 + Sqrt[3])]), (
```

```
      1 + Sqrt[3])/Sqrt[
     2 (2 + Sqrt[3])]], -((1 + Sqrt[3])/Sqrt[2 (2 + Sqrt[3])])}, {(
      1 + Sqrt[3])/Sqrt[
     2 (2 + Sqrt[3])]], -((1 + Sqrt[3])/Sqrt[2 (2 + Sqrt[3])]), -((
      1 + Sqrt[3])/Sqrt[2 (2 + Sqrt[3])])}, {-1, -1, 1}, {1, 1,
      1}}]}, Boxed -> False],
 ParametricPlot3D[{{-1 +
     2 p, (-1 + 2 p) (-1 +
       2 (1/2 - Sqrt[-(((-1 + p) p)/(2 - 4 p + 4 p^2))])), -1 +
     2 (1/2 - Sqrt[-(((-1 + p) p)/(2 - 4 p + 4 p^2))])}, {-1 +
     2 p, (-1 + 2 p) (-1 +
       2 (1/2 + Sqrt[-(((-1 + p) p)/(2 - 4 p + 4 p^2))])), -1 +
     2 (1/2 + Sqrt[-(((-1 + p) p)/(2 - 4 p + 4 p^2))])}}, {p, 0, 1},
  PlotStyle -> Directive[Thickness[.01], Black]],
 ParametricPlot3D[{-1 + 2 p, (-1 + 2 p) (-1 + 2 q), -1 + 2 q}, {p, 0,
   1}, {q, 1/2 - Sqrt[-(((-1 + p) p)/(2 - 4 p + 4 p^2))],
   1/2 + Sqrt[-(((-1 + p) p)/(2 - 4 p + 4 p^2))]}, Mesh -> None,
  PlotStyle -> Yellow]
 ]
```

### Eq.(21)

```
(*Make sure the Wolfram quantum paclet is already installed*)
QuantumBasis["Wootters"]["Matrix"] // MatrixForm
```

### Eq.(22)

```
(*Make sure the Wolfram quantum paclet is already installed*)
ArrayReshape[
 QuantumPhaseSpaceTransform[QuantumState["BlochVector"[{x, y, z}]],
   "Wootters"]["AmplitudesList"], {2, 2}]
```

### Eq.(23)

```
Solve[Simplify[
   Norm[QuantumWeylTransform[
      QuantumState[Flatten[KroneckerProduct[{p, 1 - p}, {q, 1 - q}]],
       "Wootters"]]["BlochVector"]] == 1], q, Reals] // FullSimplify
```

### Table 1

```
\[Psi]i =
  QuantumState[
   "BlochVector"[FromSphericalCoordinates[{1, \[Theta], \[Phi]}]]];
GraphicsGrid[
 Partition[
  With[{bloch =
      QuantumChannel[{StringDelete[#, "\n"], .35}][\[Psi]i][
       "BlochCartesianCoordinates"]},
     Show[QuantumState["UniformMixture"]["BlochPlot",
```

```
       "ShowLabels" -> False],
     ParametricPlot3D[
      bloch, {\[Theta], 0, \[Pi]}, {\[Phi], 0, 2 \[Pi]},
      PlotRange -> {{-1, 1}, {-1, 1}, {-1, 1}},
      PlotStyle -> Opacity[.5], Mesh -> None],
     PlotLabel -> #]] & /@ {"BitFlip", "PhaseFlip", "PhaseDamping",
    "BitPhaseFlip", "Depolarizing", "AmplitudeDamping"},
   UpTo[3]], ImageSize -> Full
 ]
```

### Table 2

```
\[Xi] = 1/3;
state = QuantumWeylTransform[
    QuantumState[Flatten[KroneckerProduct[{p, 1 - p}, {q, 1 - q}]],
      basis, "Parameters" -> {p, q}]];
sol = Normal@Solve[state["Eigenvalues"][[1]] == 0, q, Reals];
prange = Splice[(Flatten[sol // Values] /. p -> q)];
blochpotato =
  Simplify@
   QuantumWeylTransform[
     QuantumState[Flatten[KroneckerProduct[{p, 1 - p}, {q, 1 - q}]],
       basis]]["BlochVector"];
potato = ParametricPlot3D[blochpotato,
   {q, 0, 1}, {p, prange},
   PlotPoints -> 120, PlotStyle -> Opacity[.75], Mesh -> None,
   PlotRange -> {{-1, 1}, {-1, 1}, {-1, 1}}, Axes -> False,
   Boxed -> False];
bloch = QuantumState["UniformMixture"]["BlochPlot",
   "ShowLabels" -> False, "ShowAxes" -> False];
opts = {PlotRange -> {{-1, 1}, {-1, 1}, {-1, 1}}, Boxed -> False,
   Axes -> False, ImagePadding -> -40};
GraphicsGrid[Partition[(Show[ParametricPlot3D[
       Evaluate@
        FullSimplify[(QuantumChannel[#[\[Xi]]]@
            QuantumState["BlochVector"[blochpotato]])["BlochVector"],
         0 < \[Xi] < 1],
       {q, 0, 1}, {p, prange},
       PlotStyle -> Green, PlotPoints -> 50], potato, bloch,
      PlotLabel -> #, opts])
    & /@ {"BitFlip", "PhaseFlip", "PhaseDamping", "BitPhaseFlip",
     "Depolarizing", "AmplitudeDamping"}, UpTo[3]], Spacings -> -40]
```

### Eq.(32)

```
P = {p, 1 - p};
Q = {q, 1 - q} /. q -> 1/2 - Sqrt[-(((-1 + p) p)/(2 - 4 p + 4 p^2))];
ℒ1 = {{x, -x}, {-x, x}} /.
```

```
    Solve[Thread[{{x, -x}, {-x, x}} . P == D[P, p], x], x][[1]]
ℒ2 = {{y, -y}, {-y, y}} /.
    Solve[Thread[{{y, -y}, {-y, y}} . Q == D[Q, p], y], y][[1]]
ℒ =
 MatrixLog[
    KroneckerProduct[MatrixExp[ℒ1],
      MatrixExp[ℒ2]]] // ComplexExpand // FullSimplify
```

**Figure 5**

```
proba = Flatten[KroneckerProduct[{p, 1 - p}, {q, 1 - q}]];
state = QuantumWeylTransform[
    QuantumState[proba, "Wootters", "Parameters" -> {p, q}]];
sol = Solve[Simplify[Norm[state["BlochVector"]] == 1], q, Reals] //
    Simplify // Normal;

L1 = QuantumOperator[{{2/Sqrt[1 - 2 p], 0}, { 0 , 0}}];
L2 = QuantumOperator[{{0, Sqrt[-1 + 2 p]/(
    2 Sqrt[(-1 + p) p (1 + 2 (-1 + p) p)])}, { Sqrt[-1 + 2 p]/(
    2 Sqrt[(-1 + p) p (1 + 2 (-1 + p) p)]), 0}}];
\[Gamma] = (-1)^{Boole[p > 1/2], Boole[p <= 1/2]};

p0 = 0.0001;
p1 = 1;
init1 = state[<|p -> p, sol[[1]]|> /. p -> p0];
init2 = state[<|p -> p, sol[[2]]|> /. p -> p0];
final1 =
  Quiet@QuantumEvolve[
    QuantumOperator["Hamiltonian"][{L1, L2}, \[Gamma]]],
    init1, {p, p0, p1}];
final2 =
  Quiet@QuantumEvolve[
    QuantumOperator["Hamiltonian"][{L1, L2}, \[Gamma]]],
    init2, {p, p0, p1}];
Show[QuantumState["-"]["BlochPlot", "ShowAxes" -> False,
  "ShowArrow" -> False],
 Graphics3D[{Red, Thick, Arrow[{{0, 0, 0}, init1["BlochVector"]}],
   Blue, Arrow[{{0, 0, 0}, init2["BlochVector"]}]}],
 ParametricPlot3D[
  Evaluate[{final1["BlochVector"], final2["BlochVector"]}], {p, p0,
   p1}, PlotStyle -> {Directive[Thickness[.01], Red],
    Directive[Thickness[.01], Blue]}],
 ParametricPlot3D[{-1 + 2 p, (-1 + 2 p) (-1 + 2 q), -1 + 2 q}, {p, 0,
   1}, {q, 1/2 - Sqrt[-(((-1 + p) p)/(2 - 4 p + 4 p^2))],
   1/2 + Sqrt[-(((-1 + p) p)/(2 - 4 p + 4 p^2))]}, Mesh -> None,
  PlotStyle -> Yellow]]
```

**Eq.(31)**

```
entropy =
 FullSimplify[
  Total[-# Log2[#]] &@({p q,
     p (1 - q), (1 - p) q, (1 - p) (1 - q)} /. {q ->
       1/2 - Sqrt[-(((-1 + p) p)/(2 - 4 p + 4 p^2))]}), 0 < p < 1]
NMaximize[{Re@entropy, 0 < p < 1}, p]
```